# Intrinsic Localized Lattice Modes and Thermal Transport: Potential Application in a Thermal Rectifier

Michael E. Manley
Condensed Matter and Materials Division, Lawrence Livermore National Laboratory, Livermore, CA 94551, U.S.A.

**ABSTRACT**

Recent experiments provide evidence of intrinsic localized modes (ILMs) in the lattice dynamics of conventional 3D materials. Here evidence that ILMs in uranium metal enhance the thermal conductivity is presented along with speculation on how thermal transport by ILMs might be used to improve a reported design for a solid-state thermal rectifier.

**INTRODUCTION**

In materials the occurrence of nonlinear forces between atoms, or anharmonicity, causes the vibrational frequencies of the lattice vibrations to depend on amplitude. A well known manifestation is the decrease in phonon frequencies or lattice "softening" with increasing temperature. The softening of the lattice results in excess vibrational entropy which tends to stabilize an increased volume with increasing temperature and this can be associated with the well-known anharmonic thermal expansion effect [1]. Another manifestation is the phonon-phonon scattering processes that are responsible for the finite thermal conductivities of solids [1]. Both of these effects can be understood in the context of perturbation theory, where the phonons are treated as quasi-harmonic modes with softening and scattering (or lifetime) effects added on. Beyond perturbation effects, strong nonlinearity in a lattice can also give rise to new types of spatially localized vibrational modes, called an intrinsically localized modes (ILMs) [2] or discrete breathers [3]. These modes form when large-amplitude local fluctuations develop frequencies that do not resonate with the normal modes, trapping energy in dynamic "*hotspots*". At high temperatures these ILMs are expected to form randomly on the lattice stabilized by configurational entropy, much like vacancies [2].

Although known about for more than 20 years [1] only recently has evidence of the existence of ILMs in conventional 3D materials emerged. In particular, using inelastic neutron and x-ray scattering new thermally activated localized modes have been observed forming at elevated temperatures in a metallic crystal, uranium [4], and an ionic crystal, sodium iodide [5]. Experiments have also created non-equilibrium ILMs from the ground state in cold (RT) crystals [5, 6]. These experiments demonstrate that these localized modes are created by amplitude fluctuations that resemble the modes themselves, which demonstrates the nonlinear formation mechanism of an ILM [6]. Supporting data also shows that these ILMs strongly influence a surprisingly wide variety of properties; including heat capacity [4], thermal conductivity [6], thermal expansion [7], and mechanical deformation [7]. Furthermore, evidence suggests that ILMs in uranium act as an incipient driver for a solid-state phase transition [8]. Here we focus on the influence of ILMs on thermal conductivity and how this concept of dynamic nonlinear localization might be combined with a model for a thermal rectifier device [9] to produce an enhanced effect. Dynamic nonlinear localization, which tends to enhance nonlinearity by focusing energy, might also be adapted to improve other existing models for heat control devices; including thermal transistors [10], logic operations with phonons [11], and thermal memory [12].

**EXPERIMENTAL EVIDENCE**

The connection between thermal conductivity and the thermal activation of ILMs in uranium metal can be inferred by comparing the temperature where the ILMs appear in the inelastic neutron scattering data with known measurements of the thermal and electrical transport properties. Figure 1a shows the abrupt appearance of the ILM in the inelastic neutron scattering measurements somewhere between 450 K and 505 K, after Ref. [6]. Associated anomalies in the lattice dynamics, including the loss of intensity in the longitudinal optical mode along [001], indicate a "transition" to the ILM state initiates closer to 450 K [4]. In the classical theory of ILMs the activation process is expected to follow a simple exponential [2], not an abrupt transition. A rather abrupt appearance of the ILM in NaI was also observed and this activation process is still not fully understood [5]. Nevertheless, it does provide a well defined temperature to look for physical property changes and, as discussed below, this abrupt appearance may prove useful in developing thermal energy manipulating devices. Figure 1b shows the temperature dependence of the thermal conductivity and electrical resistivity; data taken from Ref. [13]. Across the temperature where the ILMs activate the electrical resistivity shows no detectable effect, bottom panel Fig. 1b. The thermal conductivity, on the other hand, shows a clear increase in its temperature dependence, top panel Fig. 1b. The lack of correspondence between the electrical and thermal transport is a violation of the Wiedemann-Franz law [1] and indicates that this change in the thermal conductivity is not due to a change in the resistance to the flow of electrons, but rather originates in the lattice vibrations. The broader implication of these results is that the ILMs act more as an energy transmission channel than as a source of scattering for energy carriers (phonons or electrons).

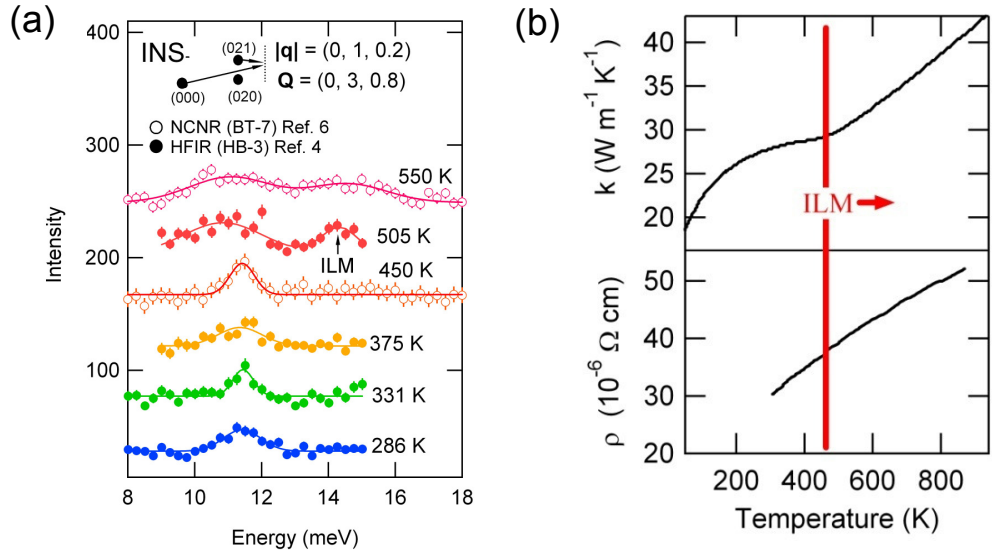

**Figure 1.** Formation of an ILM in uranium metal and associated changes in the transport properties. (a) Inelastic neutron scattering spectra on single crystal α-uranium, after Refs. [4] and [6]. (b) Transport data collected on polycrystalline α-uranium, taken from Ref. [13] ($k$ is the thermal conductivity and $\rho$ is the electrical resistivity).

**DISCUSSION**

The possibility that nonlinear lattice modes such as ILMs or other soliton-like modes could be effective carriers of heat has been considered theoretically for model systems [14]. In addition, the long-range propagation of energy by ILMs along specific crystallographic directions at specific velocities has been demonstrated for a two-dimensional hexagonal plane [15]. This effect may have some relevance here since the ILMs in uranium exist on what amounts to a nearly-hexagonal crystallographic plane [8].

The transport of heat by ILMs has some interesting ramifications for the design of devices for manipulating thermal energy. In particular, ILMs are formed by a process that spontaneously localizes energy, resulting in a significant enhancement of the usual temperature dependence of phonon frequencies. As shown in Fig. 1a the ILM "band" at 505 K appears at a frequency that is unoccupied at only slightly lower temperatures. This is an expected property of ILMs since they are expected to form in gaps or above the spectral cutoff [2]. For quasiharmoc phonons, on the other hand, it takes a large temperature change (energy input) to get a spectra feature to appear at a frequency that is significantly shifted compared to the phonon band width (several meV). In the design of the thermal rectifier proposed by Terraneo, Peyrard, and Casati [9] a frequency shift that exceeds the phonon band widths is essential for practical operation. Figure 2a shows the conceptual picture of their model for a thermal rectifier. To operate effectively, the phonon band in the middle nonlinear material must shift its frequency by enough to match the bands of the "soft" and "hard" materials on both sides. Given that normal phonon bands are typically spread over a significant frequency range means that to make this work in practice the thermal gradient has to be large enough to produce such shifts. This is especially true of efficient heat carrying bands since the necessary large group velocity implies a significant dispersion in the band, i.e. a wide band. This posses a significant practical problem since the temperature gradients would need to be 100's to 1000's of K. Heat carrying ILMs offer a simple alternative since energy is focused to the equivalent of 1000's of K locally with a heating of just a few 10's of K [2, 4, 6]. These high energies for the ILM are stabilized by configurational entropy [2]. Figure 2b indicates how an ILM-based thermal rectifier based on the same principles would function. A heat carrying band of ILMs would drop out of a phonon band providing a link to a specific band in a joined material.

Of course, this design of an ILM based thermal rectifier remains speculative since it is still not fully understood how the ILMs in uranium carry heat and whether or not this particular ILM property is unique to uranium or a more general characteristic. Nevertheless, it seems clear that thermal logic devices that make use of the nonlinear properties of lattice vibrations stand to gain a lot if they can be designed to incorporate the energy focusing properties of ILMs.

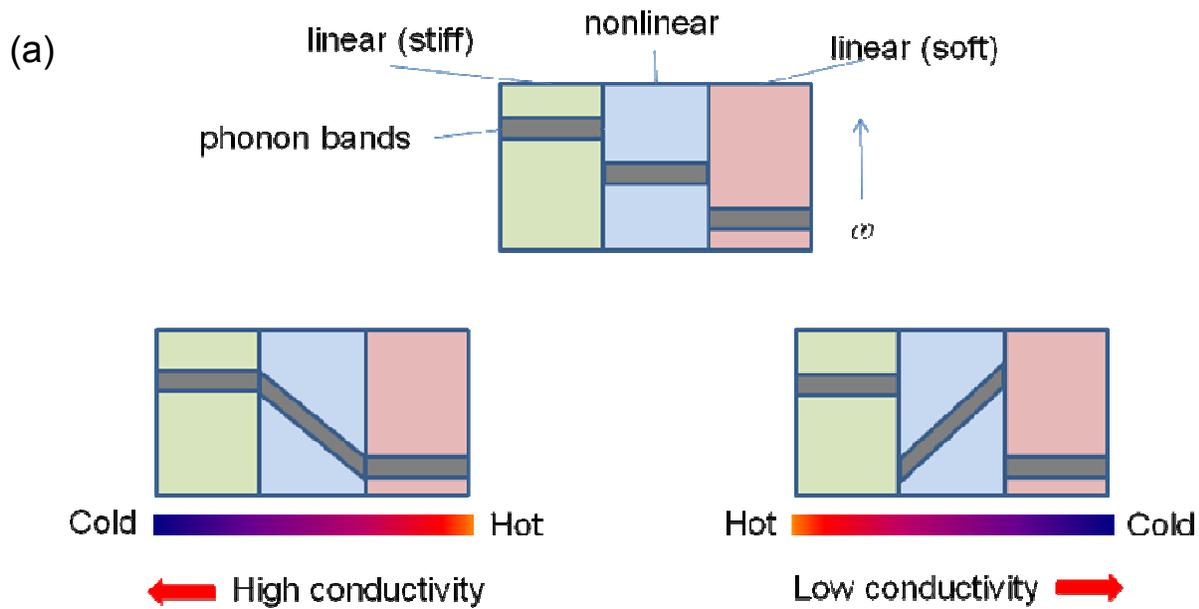

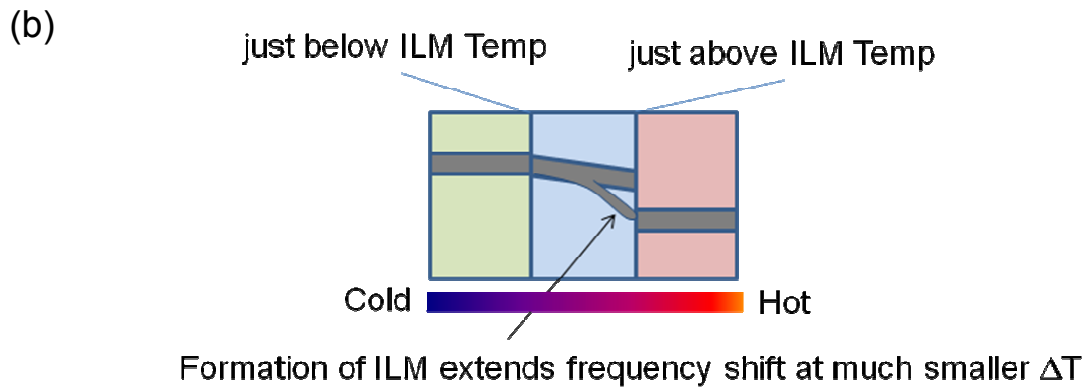

**Figure 2.** Conceptual picture of how heat carrying ILMs might be used to modify the thermal rectifier model of Terraneo, Peyrard, and Casati [9]. (a) Conceptual picture of the original model after Ref. [9]. Three materials are combined, a stiff linear material with a high frequency phonon band, a soft nonlinear material with a low frequency band, and a nonlinear material with an intermediate frequency phonon band sandwiched in between. Depending on the direction of the thermal gradient the nonlinear material produces either frequency matching at the boundaries (left) or frequency mismatching (right). The frequency matching provides high thermal conductivity and the mismatch results in low thermal conductivity. (b) A variation on this model where a band of ILMs is used to produce a large frequency shift with a small thermal gradient around the ILM activation temperature.

## CONCLUSIONS

Intrinsic localized modes, or nonlinear hotspots, have been shown to form at high temperatures in at least two classes of materials, and likely occur in all classes since they follow simply from a combination of nonlinearity and discreteness [2, 3]. They also appear to influence many materials properties including thermal conductivity. In the design of devices that specifically take advantage of nonlinearity in the lattice dynamics to control heat flow, ILMs provide an opportunity to enhance those effects. One possibility might be an efficient thermal rectifier that operates around the ILM formation temperature.